\documentclass[sigconf]{acmart}

\AtBeginDocument{
  }

\renewcommand\footnotetextcopyrightpermission[1]{}

\usepackage{enumitem}
\usepackage{adjustbox}
\usepackage{tabularx}
\usepackage{verbatim}
\usepackage{fancyvrb}
\usepackage{fvextra}  % Gives enhanced Verbatim features like wrapping
\DefineVerbatimEnvironment{Highlighting}{Verbatim}{breaklines,commandchars=\\\{\}}

\begin{document}

\title{Schemex: Discovering Structural Abstractions from Examples }

\author{Sitong Wang}
\affiliation{
  \institution{Columbia University}
  \city{\normalsize New York}
  \state{\normalsize NY}
  \country{\normalsize USA}
}
\email{sw3504@columbia.edu}

\author{Samia Menon}
\affiliation{
  \institution{University of California, Berkeley}
  \city{\normalsize Berkeley}
  \state{\normalsize CA}
  \country{\normalsize USA}
}
\email{samia_menon@berkeley.edu}

\author{Dingzeyu Li}
\affiliation{
  \institution{Adobe Research}
  \city{\normalsize Seattle}
  \state{\normalsize WA}
  \country{\normalsize USA}
}
\email{dinli@adobe.com}

\author{Xiaojuan Ma}
\affiliation{
  \institution{Hong Kong University of Science and Technology}
  \city{\normalsize Hong Kong}
  \country{\normalsize China}
}
\email{mxj@cse.ust.hk}

\author{Richard Zemel}
\affiliation{
  \institution{Columbia University}
  \city{\normalsize New York}
  \state{\normalsize NY}
  \country{\normalsize USA}
}
\email{zemel@cs.columbia.edu}

\author{Lydia B. Chilton}
\affiliation{
  \institution{Columbia University}
  \city{\normalsize New York}
  \state{\normalsize NY}
  \country{\normalsize USA}
}
\email{chilton@cs.columbia.edu}

\begin{abstract}
Creative and communicative work is often underpinned by implicit structures, such as the Hero's Journey in storytelling, design patterns in software, or chord progressions in music.
People often learn these structures from examples - a process known as schema induction. 
However, because schemas are abstract and implicit, they are difficult to discover: shared structural patterns are obscured by surface-level variation, and balancing generality with specificity is challenging.
We present Schemex, an interactive AI workflow that systematically supports schema induction by decomposing it into three tractable stages: clustering examples, abstracting candidate schemas, and contrastively refining them by generating new instances and comparing against originals. 
Studies show that Schemex produces more actionable schemas than a frontier baseline without sacrificing generalizability, with participants uncovering deep and nuanced  structural patterns.
We also discuss design implications for the cognitive role of interactive process in structure discovery.
\end{abstract}

%\begin{CCSXML}
%<ccs2012>
   %<concept>
    %<concept_id>10003120.10003121.10003129</concept_id>
    %<concept_desc>Human-centered computing~Interactive systems and tools</concept_desc>
       %<concept_significance>500</concept_significance>
       %</concept>
 %</ccs2012>
%\end{CCSXML}
%\ccsdesc[500]{Human-centered computing~Interactive systems and tools}

%\keywords{schema induction, generative AI, contrastive refinement, example-based learning, interactive workflows}

\begin{teaserfigure}
\centering
\includegraphics[width=1\textwidth]
{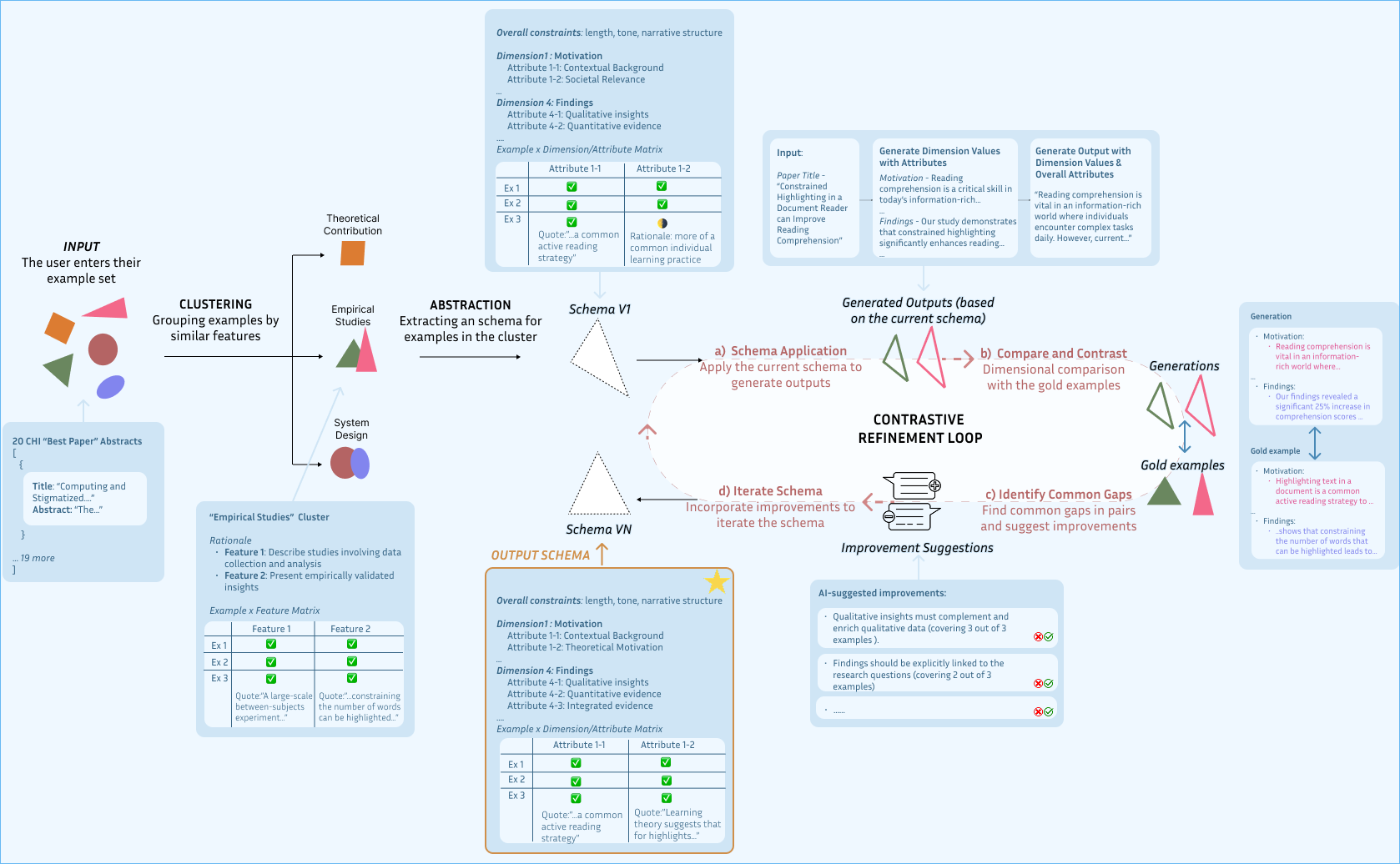}
    \caption{Schemex helps users discover structural abstractions from real-world examples in three stages: (1) clustering examples by shared structural features, displayed in a feature×example matrix; (2) abstracting a candidate schema with dimensions, attributes, and global constraints; and (3) contrastively refining the schema through a loop of applying the schema to generate outputs, comparing them against gold examples, identifying common gaps, and iterating.}
    \label{fig:teaser}
\end{teaserfigure}

\maketitle
\pagestyle{plain} 

\section{Introduction}
Beneath creative and communicative works often lies an essential unseen structure.
Stories follow recurring narrative arcs, such as the Hero's Journey~\cite{campbell2008hero}; software systems rely on shared design patterns, such as the Observer pattern~\cite{gamma1995design}; and songs with different melodies share common chord progressions~\cite{fox2013chord}.
These underlying frameworks - referred to as \textit{schemas} in cognitive science~\cite{gick1983schema} - shape how we create and understand our work.
Schemas describe the key \textit{dimensions}, their \textit{attributes}, and \textit{global constraints} that characterize the structure of instances.
For example, the Hero's Journey defines dimensions such as the Call to Adventure, the Ordeal, and the Return. Each dimension has attributes describing how it typically unfolds, such as the Call beginning when the hero is summoned to leave their ordinary world.
Global constraints like a circular narrative arc tie these dimensions together into a coherent whole.
This type of structural abstraction is powerful: it captures the essence of a wide range of instances while remaining flexible enough to accommodate variation.
While some schemas are well-known and have formal descriptions (such as those described above), many others are used without ever being explicitly identified, such as the structure of an effective system demo or a news TikTok~\cite{newman2022publishers}.
These schemas are typically learned through repeated exposure and, as a result, often remain implicit - even for experts who rely on them in practice.
Making these structures explicit can fundamentally transform how we think and create.

However, because schemas are often implicit and abstract, they are difficult to discover.
Without prior knowledge of the structure, uncovering the Hero's Journey across Star Wars, The Lord of the Rings, and The Hunger Games, for example, is far from straightforward;
it requires looking beyond surface differences in genre, plot, and setting to identify the shared dimensions, attributes, and global constraints that underlie all these stories.
Schemas must be discovered by sifting through varied data to identify abstractions that fit many instances without overfitting.
Effective schemas must balance between generality and specificity: overly abstract schemas provide little guidance for action, while overly specific ones fail to transfer to new contexts.
Furthermore, evaluating the quality of a schema is challenging in itself.
Simply inspecting a structure is insufficient; one must apply it to many instances to assess whether it is both actionable and generalizable.
This task, often referred to as schema induction, remains an open challenge.

We approach schema induction as an iterative process, similar to iterative design, where we hypothesize, evaluate, and refine schemas.
Specifically, this involves examining many examples, testing hypothesized structural abstractions against those examples, and refining them iteratively.
Because each step places significant demands on human memory and attention, we use generative AI to assist across three key stages:
\begin{itemize}\itemsep0em
\item \textbf{Clustering examples}: Not all examples share the same underlying structure (e.g., not all songs follow a vi-V-I progression). The first step is to cluster examples based on shared structures so they can be analyzed as coherent groups.
\item \textbf{Abstracting a schema from a cluster}: For each cluster, the system proposes a candidate schema capturing the key dimensions, their attributes, and global constraints shared by the examples. This schema is then mapped back onto each example to assess coherence and coverage.
\item \textbf{Contrastive refinement}: Once a candidate schema is identified, we generate new instances from it and compare them against original examples to identify mismatches. These mismatches guide schema refinement, and the cycle repeats iteratively. Unseen validation examples are included to guard against overfitting.
\end{itemize}
Throughout this process, AI accelerates data analysis and iteration, while humans remain central as reflective evaluators who shape the schema.

To operationalize these steps, we introduce Schemex (see Figure \ref{fig:teaser}), an interactive AI workflow that supports schema induction across diverse domains, from creative writing to meme creation.
Schemex guides users through the schema induction process, allowing them to interact, reflect, and revise at each stage.
Users begin by specifying a goal and providing a set of examples, for instance, a novice aiming to write HCI paper abstracts might provide a collection of CHI Best Paper abstracts.
In the clustering stage, Schemex groups examples by shared structural patterns - for instance, clustering abstracts by whether they foreground a system contribution, an empirical finding, or a theoretical argument.
For each cluster, the system presents a feature×example matrix that maps extracted features to individual examples, allowing users to assess cluster coherence and revise clusters before selecting one of interest.
In the schema abstraction stage, Schemex proposes a candidate schema for the selected cluster, consisting of dimensions, attributes, and global constraints.
For example, empirical study abstracts might yield dimensions such as Motivation, Research Questions, Study Method, and Findings, with attributes like the Findings dimension combining qualitative and quantitative evidence, and global constraints such as length and tone.
Users can inspect, add, rename, or remove any element of the schema, and evaluate consistency across examples using the matrices.
In the contrastive refinement stage, the system generates new instances from the schema given an input (e.g., a paper title) and compares them against held-out validation examples.
Users compare generated outputs with originals side by side, with text color-coded by schema dimension to support comparison.
The system proposes schema refinements based on observed mismatches, indicating how consistently each mismatch appears across examples - for example, identifying that Findings should be explicitly linked back to Research Questions in 90\% of cases.
Users can accept, reject, or revise these suggestions, iterating through an apply-test-refine cycle until they arrive at a coherent and actionable schema.

This paper makes the following contributions:
\begin{itemize}\itemsep0em
\item A framework that systematically supports schema induction, decomposing an otherwise open-ended challenge into three tractable stages: clustering, abstraction, and contrastive refinement;
\item Schemex, an interactive visual workflow that integrates AI reasoning with example-based evidence, enabling users to inspect, apply, and refine schemas through linked visualizations;
\item A technical evaluation with 12 domain experts across tasks, showing that Schemex significantly improves schema actionability (granularity, example fit, and generation quality) over o1-pro while maintaining comparable generalizability;
\item A user study with 12 participants across self-chosen tasks, with findings that even experienced practitioners uncovered structural patterns they had never explicitly articulated and design implications for the cognitive role of interactive process in structure discovery.
\end{itemize}

\section{Related Work}
 
\subsection{Schema Induction and Cognitive Foundations}
 
Cognitive science research on analogy and structural alignment shows that comparing multiple examples enables people to abstract their common structure~\cite{gick1983schema, gentner1997structure}.
In classic studies, Gick and Holyoak~\cite{gick1983schema} demonstrated that examining two analogs leads to the induction of a \textit{schema} - a generalized framework that preserves shared relations while discarding surface differences - enabling transfer to novel problems.
An effective schema abstracts away surface details to highlight underlying structure: for example, prior work on schema induction~\cite{yu2014distributed} notes that capturing an invention's purpose and mechanism is more useful than recording vague notions or overly specific traits.
 
Computational models have sought to replicate this capacity.
Kemp and Tenenbaum's hierarchical Bayesian model treats structure induction as model selection over graph-based representations~\cite{kemp2008discovery}, while DORA discovers relational concepts through analogy and represents them as explicit predicates~\cite{doumas2008theory}.
However, such models typically lack grounding in real-world artifacts, limiting practical applicability.
Schemex builds on these foundations by enabling users to perform schema induction interactively - leveraging human pattern recognition alongside algorithmic assistance to extract structural abstractions from examples.

\subsection{Interactive Abstraction and Contrastive Refinement}
 
Interactive systems increasingly support users in forming abstractions from complex data by integrating direct manipulation with automated inference.
Early work on mixed-initiative clustering~\cite{huang2010mixed} showed how adaptive grouping can incorporate user feedback, while visual interfaces like Cyclone~\cite{duman2009adaptive} and semantic interaction systems~\cite{endert2012semantic} let users reposition, merge, or split clusters to guide models toward more meaningful representations.
 
A complementary strategy uses contrast to sharpen abstractions.
Mocha~\cite{gebreegziabher2024supporting} generates counterexamples to help users refine concept definitions, and Flash Fill~\cite{gulwani2011automating} lets users correct inferred transformation rules via contrasting input-output pairs.
These systems show how surfacing edge cases accelerates convergence toward accurate models.
Schemex builds on both abstraction and contrastive techniques but extends them to schema-level generalization.
Unlike systems like Mocha that use counterexamples to refine concept labels, Schemex uses contrast to directly refine the underlying structure - comparing generated instances against real examples along dimensions and turning observed mismatches into actionable edits.

\subsection{HCI Systems for Structure Discovery}
 
In design and content domains, many systems surface reusable patterns by mining common structures at scale.
Webzeitgeist mined UI structures from thousands of websites~\cite{kumar2013webzeitgeist}, and other work has identified design patterns across mobile apps through large-scale clustering~\cite{alharbi2015collect}.
Prior work has also emphasized documenting reusable solutions in structured formats such as problem-context-solution~\cite{kruschitz2010human}.
For multimodal content, RecipeScape~\cite{chang2018recipescape} clusters recipes by procedural similarity, hierarchical tutorial generators~\cite{truong2021automatic} decompose videos into stepwise schemas, and VideoMix~\cite{yang2025videomix} aggregates how-to content into coherent workflows.
 
Sensemaking tools help users organize data and externalize evolving structures~\cite{gero2024supporting}.
Jigsaw~\cite{stasko2007jigsaw} supports document analysis through interactive visualizations, Selenite~\cite{liu2024selenite} provides corpus overviews with LLM assistance, and CollabCoder~\cite{gao2024collabcoder} supports team-based qualitative coding.
LLooM~\cite{lam2024concept} enables concept induction by helping users cluster and label high-level ideas from unstructured text.
Crowdsourced approaches further show that externalizing schemas can improve ideation, analogy retrieval, and collaborative understanding~\cite{yu2014distributed, yu2014searching, kittur2014standing}.
 
Schemex shares the goal of supporting structure discovery but differs in \textit{what} is being discovered.
Tools like LLooM and CollabCoder perform \textit{concept induction} - identifying \textit{what} examples are about by producing topical labels, categories, or codes.
Schema induction, by contrast, asks \textit{how} examples are internally organized: it seeks the underlying dimensions, attributes, and constraints that characterize their shared structure.
Because schemas capture structure rather than topic, they are \textit{generative} - a schema for CHI abstracts can produce new abstracts for unseen papers, whereas a set of thematic codes cannot.
Schemex also differs from tools like DesignWeaver~\cite{tao2025designweaver} and Luminate~\cite{suh2024luminate}, which analyze AI outputs from known prompts; Schemex addresses the harder problem of discovering new, verified structural knowledge from messy, real-world examples that extend beyond what LLMs already encode.

\subsection{Schemas Used in HCI Systems}
 
Because of the generalized scaffolding they provide - particularly for novices - schemas are widely used in creativity support tools.
Books of design patterns have influenced generations of architects~\cite{patternlanguage}, software engineers~\cite{gamma1995design}, and web designers~\cite{designofsites}.
In computer graphics, design patterns help solve challenges such as furniture layout~\cite{interior_design}, map generation~\cite{line_drive}, and illustrating assembly instructions~\cite{maneeshdesignprinciples}.
In HCI, design patterns have supported filmmakers~\cite{motif,leake_video_editing}, human-robot interaction~\cite{design_patterns_prototyping_HRI, design_patterns_social_HRI}, and product ideation~\cite{yu2014distributed, yu2014searching}.
 
AI has also been instrumental in helping users apply abstract schemas to creative tasks such as writing stories~\cite{dramatron}, creating visual blends~\cite{Chilton2019VisiBlends,Chilton2021VisiFit,Cunha2018ShellHornUnicorn,Wang2023PopBlends}, producing teasers~\cite{Wang2024PodReels}, crafting social media videos~\cite{Wang2024ReelFramer, menon2024moodsmith}, and generating inclusive emergency messages~\cite{Jit2024WritingStorm}.
In nearly all of these systems, the schemas were constructed manually - a labor-intensive process.
Accelerating schema induction could vastly expand access to structured creativity, enabling more people to design, communicate, and express ideas effectively.
\section{Schemex System}
 
\begin{figure*}
\centering
\includegraphics[width=0.94\textwidth]{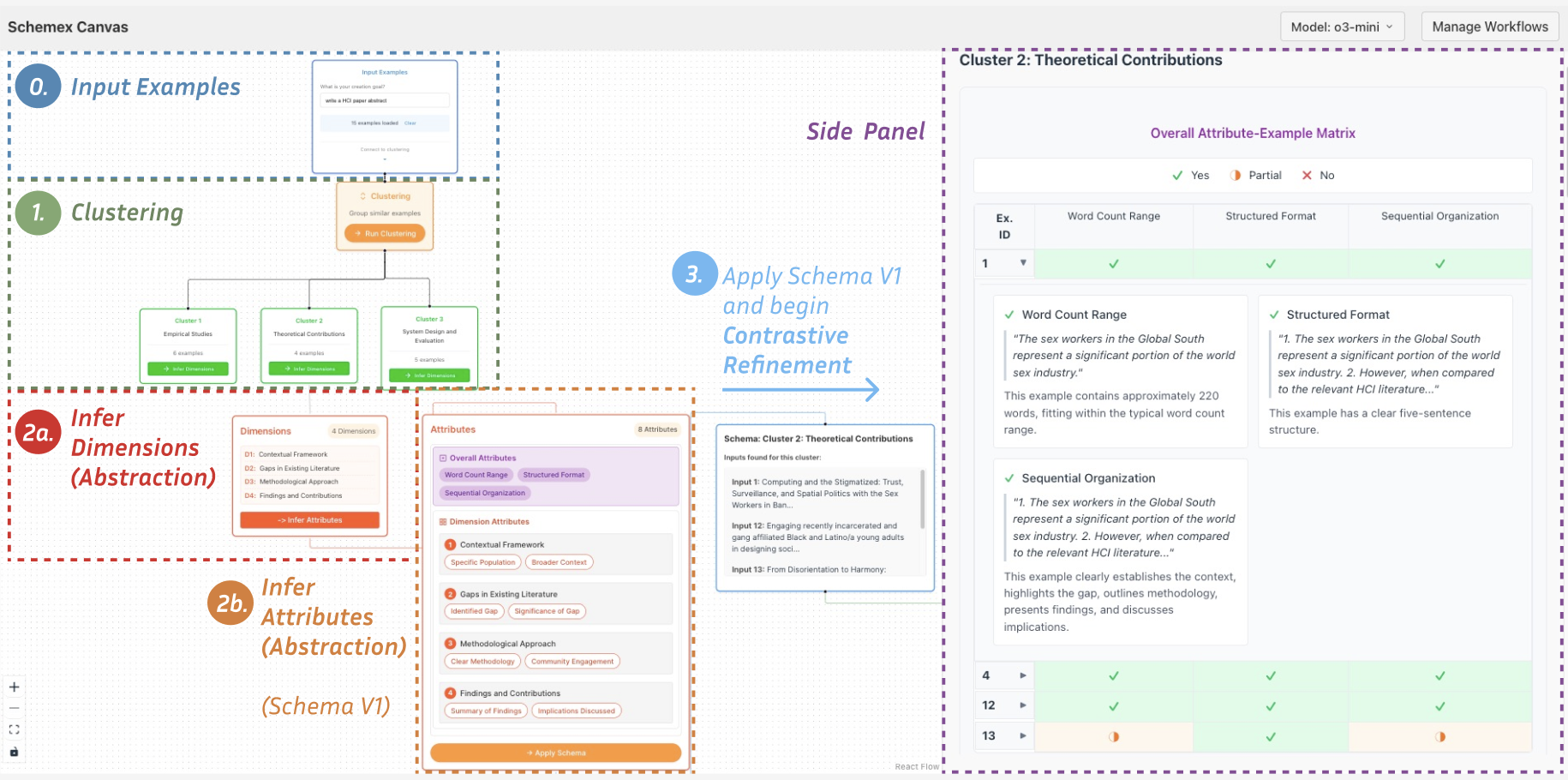}
\caption{Schemex guides users through schema induction given their input examples: (1) clustering, (2a--b) abstraction, and (3) contrastive refinement. Each node opens a side panel (right) with definitions, example mapping matrices, and inline citations.}
\label{fig:system_screenshot}
\end{figure*}
 
\subsection{Design Goals}
 
Schemex was developed to help users infer underlying schemas from messy real-world examples, supporting deeper understanding and more effective authoring. 
Its design is guided by four core goals:
 
\begin{enumerate}[nosep,leftmargin=*,label={\textit{DG{\arabic*}}}]
\item \textbf{Ensure schemas remain grounded in evidence.}
Schema induction must remain accountable to the source examples. Schemex grounds both AI-generated proposals and human judgments in evidence through example matrices, explicit rationales, and inline citations.
 
\item \textbf{Ensure examples share structural similarities before developing schemas.}
Generalizing across structurally different examples can result in overly broad schemas. Schemex addresses this by clustering examples that share latent structural similarities, preventing overgeneralization and establishing a basis for abstraction.
 
\item \textbf{Evaluate schema quality by applying it to examples.}
Schema quality is difficult to assess in the abstract - patterns that appear coherent in isolation often break down when applied across diverse cases. Schemex operationalizes this by applying schemas to examples, examining contrasts, and supporting iterative revision.
 
\item \textbf{Allow human editing of schemas.}
While AI can propose clusters and structures, it lacks contextual awareness of users' goals and domain knowledge. Effective schema induction thus requires human oversight to validate clusters, accept or reject structural elements, and guide revisions. Schemex provides lightweight, localized controls that preserve user agency while keeping iteration fast and cognitively manageable.
\end{enumerate}
 
\subsection{User Walkthrough}

Consider Sam, a first-year graduate student learning to write HCI abstracts. Sam begins by entering their learning goal - ``Write an HCI paper abstract'' - and uploads 20 CHI Best Paper abstracts along with their titles. The system stores each abstract-title pair and automatically sets aside a validation subset for later testing against overfitting.

\subsubsection{Stage 1: Clustering.}
\begin{figure}
\centering
\includegraphics[width=1\linewidth]{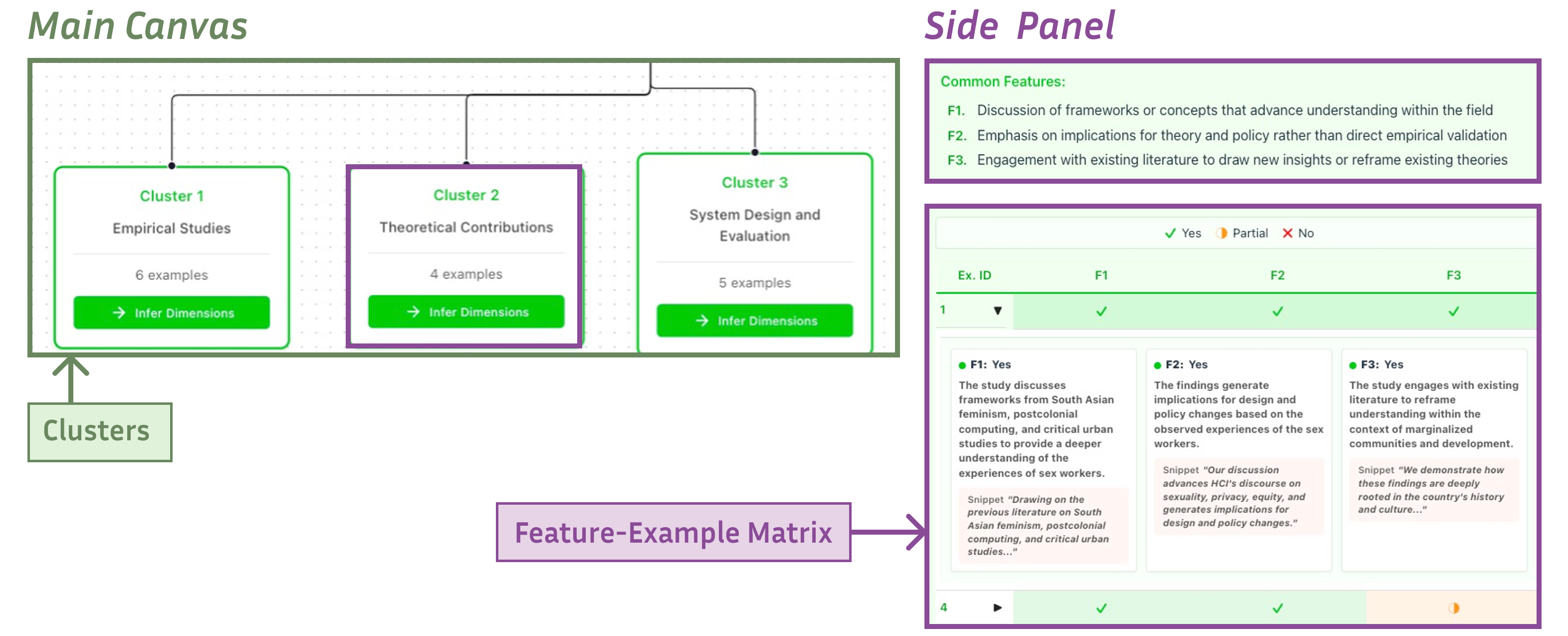}
\caption{Schemex Stage 1: Clustering.}
\label{fig:clustering}
\end{figure}

Sam clicks ``Run Clustering'' (see Figure \ref{fig:clustering}). Schemex processes all examples at once and proposes three clusters: \textit{Empirical Studies}, \textit{Theoretical Contributions}, and \textit{System Design \& Evaluation}, reflecting structurally distinct abstract types that emerged from the examples: for instance, empirical abstracts center on research questions, methodology, and findings; while system abstracts follow a build-then-validate arc. Each cluster is represented as a node on the interface. Curious about theory papers, Sam clicks on the \textit{Theoretical Contributions} node. A side panel opens, showing common features shared by this cluster: 1) \textit{Discussion of frameworks or concepts that advance understanding;} 2) \textit{Emphasis on implications for theory or policy;} 3) \textit{Engagement with existing literature to reframe or extend prior theories.}
 
Below this summary, a feature-example matrix lists each abstract in the cluster and indicates whether it exhibits each feature (yes/partial/no), with citations to the supporting text spans. 
For example, the snippet from one abstract - \textit{``...generates implications for design and policy changes''} - demonstrates alignment with the cluster's second feature.
Sam skims the side panel and confirms that the clustering is coherent.
They decide to proceed with the theory cluster.
This ensures schema induction is grounded in evidence (DG1) and prevents overgeneralization (DG2).

\subsubsection{Stage 2: Abstraction.}
 
\begin{figure}
\centering
\includegraphics[width=1\linewidth]{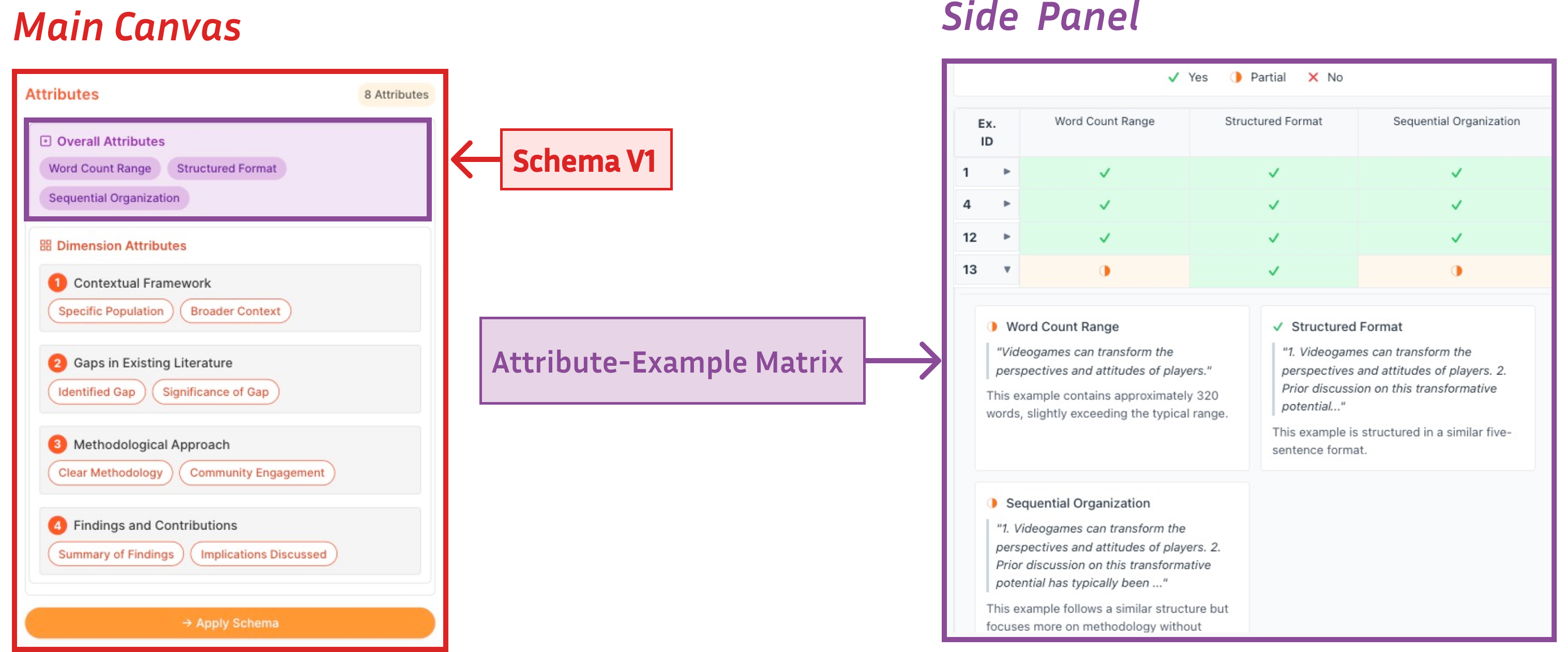}
\caption{Schemex Stage 2: Abstraction.}
\label{fig:abstraction}
\end{figure}

Sam clicks ``Infer Dimensions'' to dive deeper into the \textit{Theoretical Contributions} cluster (see Figure \ref{fig:abstraction}). 
Schemex proposes four candidate dimensions: \textit{Contextual Framework}, \textit{Gaps in Existing Literature}, \textit{Methodological Approach}, and \textit{Findings}. Sam opens the side panel to view how individual examples map to each dimension via a yes/partial/no matrix linked to citations from the example abstracts.
 
Sam then clicks ``Infer Attributes.'' The system generates dimension-specific attributes (e.g., \textit{Contextual Framework} $\rightarrow$ \textit{Specific Population}, \textit{Broader Context}; \textit{Findings} $\rightarrow$ \textit{Summary of Findings}, \textit{Implications Discussed}) and global constraints (e.g., \textit{Word Count Range}, \textit{Structured Format}, \textit{Sequential Organization}). The side panel shows another matrix mapping attributes to examples, with highlighted spans justifying each judgment.
Together, these elements form an initial schema (see Schema V1 in Figure \ref{fig:abstraction}): a structured characterization consisting of named dimensions (the core structural components), their attributes (how each is typically realized across instances), and global constraints (cross-cutting properties such as length, tone, and organizational arc).
 
Sam notices that the \textit{``Specific Population''} attribute feels too vague for their needs. They hover over the label, edit it to \textit{``Focal Community,''} and see the change instantly reflected in the interface. They leave other attributes unchanged, appreciating that the system allows lightweight edits. This process keeps abstractions tightly connected to evidence (DG1) while preserving Sam's ability to refine them interactively (DG4).

\subsubsection{Stage 3: Contrastive Refinement.}
 
\begin{figure}
\centering
\includegraphics[width=1\linewidth]{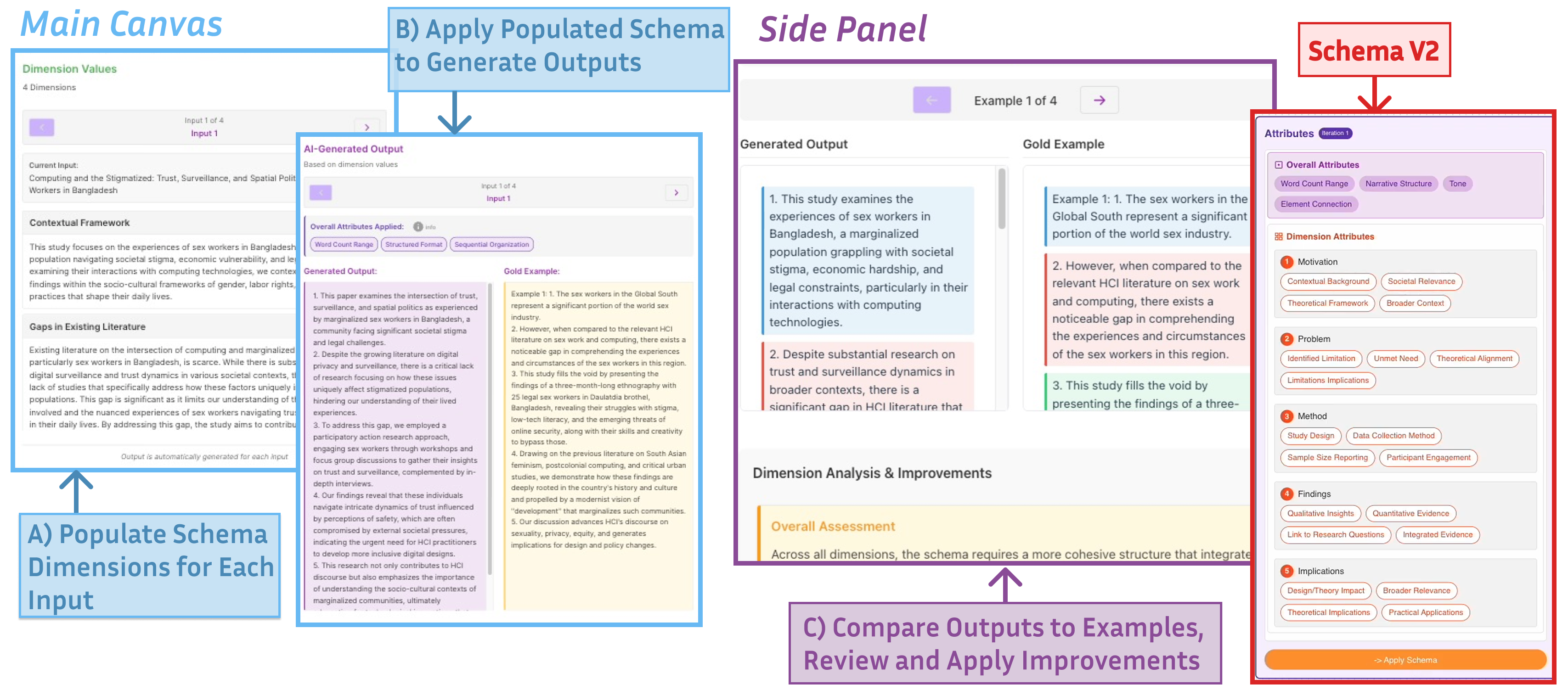}
\caption{Schemex Stage 3: Contrastive Refinement.}
\label{fig:refinement}
\end{figure}

With the schema in place, Sam clicks ``Apply Schema'' to test its effectiveness (see Figure \ref{fig:refinement}). Schemex generates per-dimension values for each paper based on its title and the attributes defined for each dimension, then composes full draft abstracts that integrate these values with the global constraints. The interface presents drafts side by side with the corresponding original abstracts, including the held-out validation examples that were set aside at the start. Sentences are color-coded by dimension, allowing Sam to align generated text with its counterpart in the original.
 
Schemex also detects gaps between generated and original examples. For instance, it flags that most generated outputs summarize findings without linking them to existing theoretical frameworks. Based on this pattern, the system suggests adding a new attribute, \textit{theoretical integration}, under the \textit{Findings} dimension. Sam reviews this proposal alongside the color-coded pairs, choosing to accept it. Sam can also reject, edit or add to these suggestions. The schema updates accordingly, and the system regenerates outputs using the refined version (see Schema V2 in Figure \ref{fig:refinement}).
 
With each cycle of applying, inspecting, and refining, the schema becomes sharper. This process makes schema quality visible through application (DG3) while keeping Sam in control (DG4). The resulting schema makes explicit what experienced writers do implicitly - giving Sam a structural map to inform and guide their own writing.

\subsection{Key Design Features}
 
\subsubsection{Multimodal preprocessing.}
To support schema induction from image and video examples, Schemex converts these artifacts into a unified textual representation using vision models for visual descriptions and automatic speech recognition for audio transcription.
 
\subsubsection{Stage 1: Clustering with rationales and matrices.}
Schemex uses a reasoning model to analyze examples and propose clusters with rationales. The cluster side panel presents common features and a yes/partial/no feature$\times$example matrix; clicking a cell reveals inline citations for that judgment. Users can revise clusterings and choose which cluster(s) to pursue.
 
\subsubsection{Stage 2: Abstraction into dimensions, attributes, and global constraints.}
Schemex proposes dimensions, attributes, and global constraints - each with definitions - and populates an example mapping matrix with yes/partial/no judgments and citations. Users can edit, add, remove, or rename schema elements, or return to clustering if the emerging schema is too coarse or narrow.

\begin{table*}[t]
\begin{adjustbox}{width=0.8\textwidth}
\centering
\begin{tabular}{|l|l|l|l|l|l|l|} \hline
 & Condition 1: & Condition 2: & Condition 3: & p-value & p-value & p-value \\
 & o1-pro & Schemex Initial & Schemex Full & (1 vs 2) & (2 vs 3) & (1 vs 3) \\ \hline \hline
Schema Granularity & 5.0 (0.85) & 5.9 (0.67) & 6.8 (0.39) & 0.026 & \textbf{0.002} & \textbf{0.003} \\ \hline
Example Fit & 5.6 (0.90) & 6.1 (0.67) & 6.6 (0.51) & 0.19 & \textbf{0.014} & \textbf{0.018} \\ \hline
Generation Quality & 5.5 (0.52) & 5.8 (0.58) & 6.3 (0.65) & 0.16 & \textbf{0.014} & \textbf{0.013} \\ \hline
Generalizability & 5.0 (1.10) & 5.5 (0.67) & 5.8 (0.75) & 0.19 & 0.26 & 0.07 \\ \hline
\end{tabular}
\end{adjustbox}
\caption{Results comparing three conditions. We report means, standard deviations, and p-values from paired-sample Wilcoxon tests, corrected for multiple comparisons using the Benjamini-Hochberg procedure. Bolded p-values are statistically significant.}
\label{tab:tech_results_merged}
\end{table*}

\subsubsection{Stage 3: Contrastive refinement through apply-and-test.}
Schemex applies the schema by producing per-dimension values and composing full drafts under the global constraints. A comparison view uses per-dimension color coding to align generated text with the original corpus and lists dimension-scoped improvement suggestions with consistency indicators. Users accept or reject suggestions, modify the schema, and reapply to observe effects.
 
\subsubsection{Interaction scaffolds.}
A node-based canvas shows the full pipeline and its artifacts; users can enter at any node to re-run or revise without losing context. A persistent side panel consolidates definitions, matrices, and inline citations so that evidence is always one click away. Localized controls - opening rationales, filtering matrices, accepting or rejecting suggestions, and editing schema elements - keep iteration fast while preserving agency.
Implementation details are provided in Appendix~\ref{appendix:implementation}.

\section{Technical Evaluation}

To evaluate the quality of schemas produced by Schemex, we conducted a controlled study in which domain experts assessed schemas generated under three conditions:
 
\begin{itemize}\itemsep0em
\item Condition 1: o1-pro - a state-of-the-art extended reasoning model 
that uses long internal chains of thought and self-correction over a 
200K-token context window.
\item Condition 2: Schemex Initial - our structured workflow with clustering followed by abstraction into a schema.
\item Condition 3: Schemex Full - the full system with one iteration of contrastive refinement building on Schemex Initial.
\end{itemize}
 
We hypothesize that schemas generated under Schemex Full will improve schema \textit{actionability} - measured through granularity, example fit, and generation quality - over the other conditions, while maintaining comparable generalizability.
 
\subsection{Data Preparation}
We collected 6 topics as our test suite, each containing 15 randomly selected examples spanning academia, business, and media across different modalities:
(1) EE/CS faculty candidate job talk abstracts,
(2) UIST teaser videos,
(3) LinkedIn profile summaries,
(4) startup pitch decks,
(5) news headlines, and
(6) news TikToks~\cite{newman2022publishers}.
Examples were collected from trustworthy real-world sources (e.g., university emails for job talk abstracts, the UIST YouTube channel for teaser videos) and processed through our multi-modal analysis pipeline into structured textual data.
 
To generate schemas with o1-pro, we used a prompt equivalent to Schemex's task description and input (see Appendix~\ref{appendix:prompts_o1_pro}).
Each condition was run once per task.
All schemas were then used with GPT-4o to generate outputs: for each schema (per cluster), we generated outputs for two randomly selected inputs (one if the cluster contained a single example).
This yielded 20 schemas and 38 outputs for Condition 1, and 17 schemas and 33 outputs for Conditions 2 and 3 accordingly.
 
\subsection{Participants and Procedure}
Twelve domain experts (two per topic; average age 27.0; 7 female, 5 male) evaluated the schemas and outputs.
Each expert had relevant domain experience.
For example, journalism graduate students with at least two years of experience making news TikToks, or university career coaches who regularly advise on LinkedIn profiles.
Before evaluation, experts reviewed example schemas and a rubric covering four dimensions (Table~\ref{tab:tech_results_merged}).
They read the original examples thoroughly, then independently rated each dimension on a 7-point scale with justifications.
Condition labels were removed and the order was counterbalanced.
Experts were compensated \$50/hour.
 
\subsection{Evaluation Results and Findings}
 
Interrater agreement (within ±1 point) was 0.78, indicating substantial agreement given the subjective nature of the task. We averaged the two experts' scores and ran paired-sample Wilcoxon tests to compare conditions, with p-values corrected for multiple comparisons using the Benjamini-Hochberg procedure. Results are reported in Table ~\ref{tab:tech_results_merged}.
 
\subsubsection{Schemex improves schema actionability over the baseline.}
 
Across all three actionability dimensions, Schemex Full significantly outperforms o1-pro: schema granularity (p=.003), example fit (p=.018), and generation quality (p=.013). Schema granularity improved progressively across the three conditions, with statistically significant gains emerging in the full pipeline. Together these results suggest that the complete Schemex pipeline is necessary to reliably improve schema actionability over a strong baseline, with contrastive refinement providing measurable gains beyond clustering and abstraction alone.
 
\paragraph{Schema Granularity.}
Schemex offers more granular, grounded schemas than o1-pro. This is partly enabled by more coherent clustering upstream. For example, in the job talk case, o1-pro produced thematic clusters ("Systems-Focused" vs. "Human Impact-Focused") where a talk on user-centered analytical interfaces could fit either, while Schemex identified a clear structural divide ("Sequentially Ordered Roadmap Abstracts" vs. "Integrated Narrative Abstracts") that enabled more precise schema extraction.
 
The resulting schemas reflect this precision. In the LinkedIn profile case, while o1-pro noted broad patterns like "often include structured bullet points" or "end with an invitation to connect," Schemex identified precise narrative moves such as a "pivotal challenge or turning point" that details career evolution - capturing how personal storytelling actually functions in the examples.
 
\paragraph{Example Fit.}
Through contrastive refinement, Schemex Full achieves finer-grained alignment between schema attributes and examples. For instance, for news headlines, Schemex Initial emphasized active verbs and single-event focus. The full version added guidance for compound action handling - such as using "and," "as," or "after" to connect multiple developments - which better reflects the examples that combine actions and consequences in a single headline.
 
\paragraph{Generation Quality.}
Schemex Full produces schemas that more effectively guide generation. For the UIST demo walkthrough cluster, while o1-pro mentioned key elements like on-screen annotations, Schemex detailed how annotations are layered to guide user understanding - making the schema more actionable for producing outputs that match the quality of original examples.
 
\subsubsection{Generalizability: an inherent tradeoff.}
Generalizability scores improved across conditions (5.0→5.5→5.8), though differences did not reach statistical significance (all p > .05). The directional trend suggests Schemex's pipeline does not come at the cost of generalizability while consistently improving actionability.

That said, the result points to a real tension inherent to evidence-based schema induction: schemas grounded in a specific example set will, by design, reflect the conventions of that corpus. Contrastive refinement sharpens schemas against real examples, improving actionability but not inherently expanding their scope. In the pitch deck task, for instance, Schemex Full emphasized storytelling patterns prominent in the example set, while o1-pro's schema focused on underlying reasoning structure regardless of narrative form, lending o1-pro's schema more flexibility for novel formats.

Whether this flexibility is desirable depends on the user's goal. For a novice learning genre conventions, a grounded and specific schema is often more immediately useful than a broadly transferable but less actionable one. The consistency indicators attached to each refinement suggestion further allow users to distinguish broadly-held patterns from minority ones, giving them deliberate control over how tightly the schema fits their example set. Incorporating more diverse examples can also broaden schema scope without sacrificing groundedness.
\section{User Study}

\begin{table}
    \centering
    \begin{tabular}{|c|c|c|} \hline 
         \textbf{Participant}& \textbf{Example set} & \textbf{Number}\\ \hline 
         P1& Personal creative writing & 10 \\ \hline 
         P2& NBA team's Instagram posts & 13\\ \hline 
         P3& Iambic Pentameter & 15 \\ \hline 
         P4& Subheaders from NPR Articles & 14 \\ \hline 
         P5& Tech Career LinkedIn Bios &  8 \\ \hline 
         P6& HCI PhD SoP First Paragraph & 12\\ \hline 
         P7& Fandom Wiki for Harry Potter & 10 \\ \hline 
         P8& Hopecore TikTok & 8 \\ \hline
         P9& Recipe blog introductions & 12 \\ \hline
         P10& GitHub README files & 10 \\ \hline
         P11& Wedding toast speeches & 9 \\ \hline
         P12& Product Hunt launch posts & 11 \\ \hline
    \end{tabular}
    \caption{Participants and the example sets they chose.}
    \label{tab:user_study_tasks}
\end{table}

We conducted a user study with twelve participants who used Schemex on self-chosen tasks to investigate the research question: \textit{How do users engage with the interactive structure discovery workflow, and how does it support their structural understanding?} Specifically, we examine how each stage contributes to this, how users shape schemas through interaction, and what design implications emerge.

\subsection{Protocol}

We recruited twelve participants (average age=26.3, six female, six male) through a university mailing list and word-of-mouth. All 
participants reported being very familiar with ChatGPT. Each participant selected an example set of their choice, spanning diverse domains and modalities, with 8-15 examples per set (see Table~\ref{tab:user_study_tasks}).

Each participant completed two tasks: (\textit{Task A}) inducing a schema using Schemex, and (\textit{Task B}) inducing a schema using  ChatGPT (o1-pro) for the same example set. We used ChatGPT (o1-pro)  as a reference condition, as it was the strongest publicly available tool for complex reasoning tasks at the time of the study. Having participants complete both conditions gave them a concrete point of comparison when reflecting on their experience during interviews. To control for ordering effects, half began with Task A and half with Task B, with a brief demo before Task A. Each task had a time limit of 30 minutes. After each task, participants completed a short questionnaire rating schema clarity, structural insight, novel reflection, confidence, and visual presentation on 1-7 scales, followed by a semi-structured interview. Each session lasted approximately 80 minutes, and participants were compensated \$25/hour.

\begin{figure}
\centering
\includegraphics[width=1\linewidth]{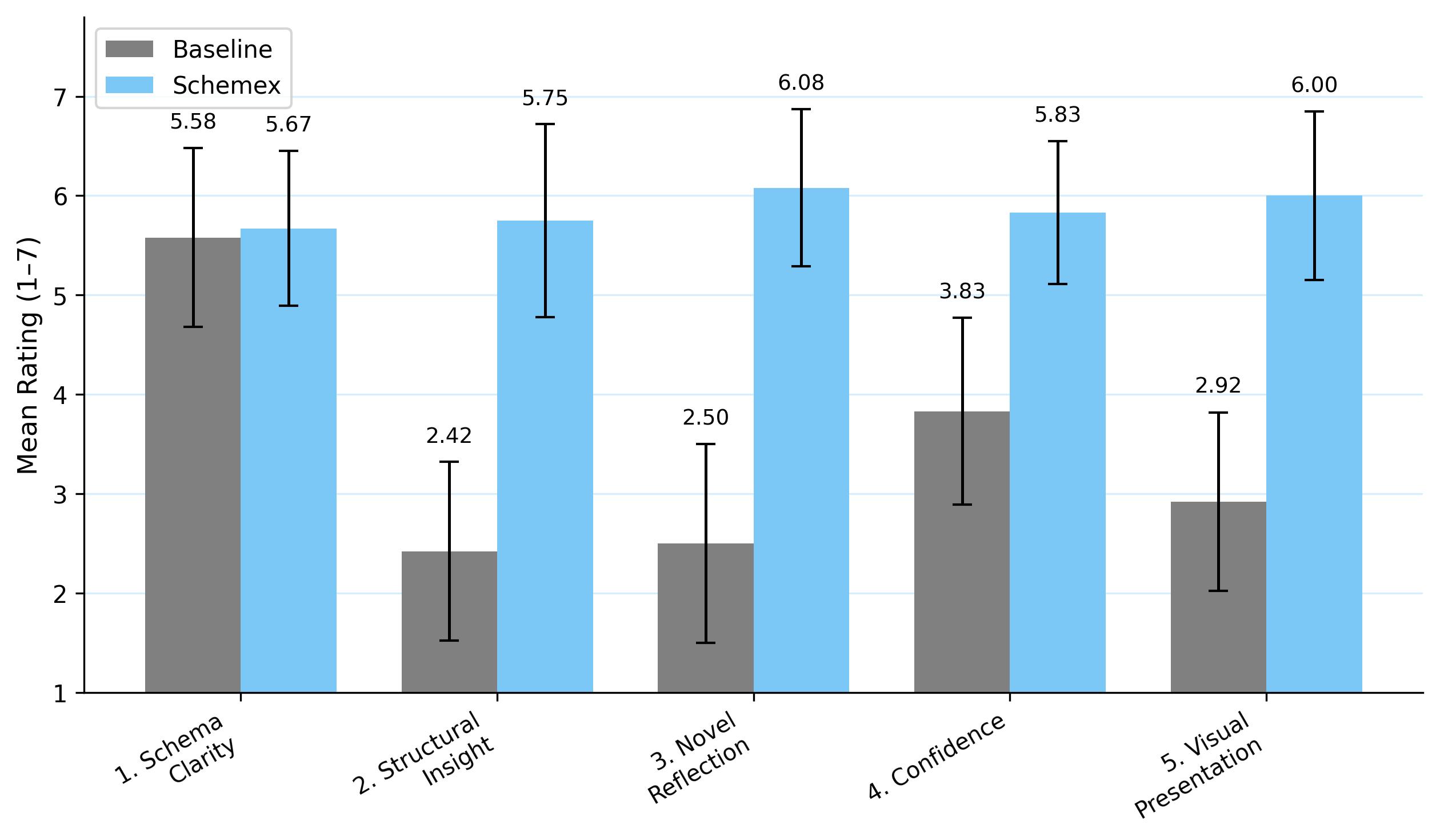}
\caption{Participant ratings over five dimensions for baseline vs. Schemex.}
\label{fig:statistics}
\end{figure}

Self-reported ratings are summarized in Figure~\ref{fig:statistics}. 
Schemex received higher ratings than ChatGPT o1-pro on structural insight, novel reflection, confidence, and visual presentation; schema clarity was comparable across conditions. Given the interface-level differences between conditions,  we treat these ratings as indicative of participants' experience rather than as performance claims. Our analysis, therefore, focuses primarily on qualitative insights from thematic analysis~\cite{braun2006using}: The first author conducted an initial coding pass, from which the research team collaboratively developed the final themes below.
 
\subsection{Findings}
Across all themes, a consistent pattern emerged: participants gained structural understanding not just from the schemas Schemex produced, but from the process of building them - with each stage of interaction surfacing insights they had not anticipated and could not have reached by simply receiving a finished schema.

\subsubsection{Clustering as a Discovery Mechanism}
 
One of the most consistent findings was that clustering surfaced structural distinctions participants had not previously recognized, even in domains where they considered themselves knowledgeable.
Many users initially believed their example sets were homogeneous; seeing examples organized into structural groups prompted re-evaluation.
 
P4, analyzing NPR subheaders, discovered that what he considered ``the same thing'' actually comprised meaningfully different categories such as ``Direct Factual Summaries'' and ``Composite and Contrasting Narratives''.
He explained: \textit{``It helps you understand something that seems like `the same thing' [and find] there's different types of them.''}
P11, analyzing wedding toasts, described a similar experience: \textit{``I didn't expect there to be structure at all - I thought people just say whatever feels right. But some people tell a timeline story, and others jump between jokes and appreciation.''}
P1, who analyzed samples of her own creative writing, expressed a different kind of surprise: the system \textit{``found the implicit connections and anxiety around time''} across pieces she had considered unrelated.

The feature-example matrices played a key role in making these groupings credible.
Rather than simply presenting cluster labels, the system showed which features were shared (or partially shared) across examples, with inline citations to specific text spans.
P2 noted that this granularity helped him \textit{``see the added context''} behind the grouping, rather than taking it on faith - while also helping participants \textit{scope} which structure to pursue before investing effort in abstraction.
 
\subsubsection{Grounding Abstractions in Evidence}
 
Across interviews, participants consistently emphasized the value of being able to trace schema elements back to specific examples.
This grounding mechanism - realized through the example mapping matrices and inline citations - shaped both comprehension and trust.
 
P1 described this as balancing abstraction with concreteness: \textit{``I love that it helped balance the abstractness... I don't think an AI-empowered interface could do without examples.''}
She contrasted this with the baseline, noting that \textit{``the way ChatGPT engaged with the examples was generally summarizing''} - producing plausible-sounding structures without making their evidential basis transparent.
P5 connected grounding directly to confidence: \textit{``I think it also goes to the trust in the system. Examples are the most helpful for me.''}
P2 emphasized the matrix as a practical evaluation tool: \textit{``It was just an easy way to sort through [the information] ... and not have to go back and figure out each one.''}
 
The matrices were particularly valuable for distinguishing genuine patterns from superficial ones.
P10, working with GitHub README files, noted: \textit{``I could actually check - does this really show up in different repos or just one? The matrix made it obvious when something was consistent versus kind of overfitting.''}
P12, analyzing Product Hunt launch posts, echoed this: \textit{``The citations helped me see when a pattern was real versus just something that sounded right.''}
 
This evidence-grounding design also changed how participants handled disagreement with the system.
Rather than accepting or rejecting proposals wholesale, users could inspect the underlying evidence and make targeted edits.
 
\subsubsection{Contrastive Refinement as a Reflection Tool}
 
The contrastive refinement stage - where the system generates new instances from the schema and compares them against originals - served as more than a quality check.
Participants described it as a mechanism for \textit{externalizing} their implicit understanding, making gaps visible that were otherwise difficult to articulate.
 
P1 highlighted how the system's suggestions scaffolded iteration: \textit{``Schemex helps you with how you should iterate and provides you with an understanding of all of the opportunities.''}
She contrasted this with the baseline: \textit{``I know what needs to change, but I don't know what to change to. [I have to] act in a very corrective way.''}
 
The color-coded dimension alignment between generated and original text helped participants quickly identify where schemas fell short.
P3, working with iambic pentameter, found that the comparison step revealed subtle patterns in rhetorical structure that neither he nor the initial schema had captured.
His final schema included specific instructions regarding meter, language style, and the formation of rhetorical questions - details that emerged through iterative refinement rather than initial abstraction.
P9 described a similar experience with recipe blog introductions: \textit{``The generated intros had the structure but were missing that sensory part - the smell, the texture. That's when I realized that needed to be its own thing in the schema.''}
P11, analyzing wedding toasts, identified a gap in emotional specificity: \textit{``The generated versions felt generic, like they could be about anyone. The real ones had callbacks to very specific shared moments - that wasn't in my schema at first.''}
 
The consistency indicators attached to each suggestion also helped participants prioritize.
When the system flagged that a mismatch appeared in most of the comparisons, participants treated it as a strong signal; when consistency was lower, they investigated further before deciding.
 
\subsubsection{Human Steering and Agency}

Throughout the workflow, participants did not passively consume system outputs - they actively shaped the schema induction process based on their own goals and domain knowledge.

At the clustering stage, participants chose which cluster to pursue based on what they wanted to learn.
P5, analyzing LinkedIn bios, selected a cluster that aligned with her professional interests rather than the largest cluster.
At the abstraction stage, participants tailored schemas to their needs.
P12, working with Product Hunt posts, described splitting a broad dimension into ``what the product does'' versus ``why it matters,'' after which \textit{``the examples mapped way more cleanly.''}
At the refinement stage, participants used the apply-compare cycle not just to evaluate the schema but to deepen their own understanding, selectively accepting suggestions and re-running generation to test their intuitions.

P2 found that this level of specificity prompted deeper thinking: \textit{``[With Schemex] I found - `this is a thing you do and here's the precise way it manifests.' It makes me reflect.''}
P3 felt the resulting schema matched what an extended manual study would yield: \textit{``It was just so much more descriptive and aligned with what I hope I would take away if I had a lot of time spent studying this.''}

Participants also valued that the system made its own reasoning transparent rather than presenting results as final.
P5 appreciated \textit{``the self-reflection of the AI,''} referring to how the comparison step surfaced the schema's limitations alongside its strengths.
P8 noted that while the interface had \textit{``a lot going on,''} it \textit{``was laid out super visually - it was easy to follow,''} suggesting that the node-based canvas helped manage complexity without sacrificing depth.
That said, not all participants preferred the rich interface: P5 expressed a preference for condensed, linear representations: \textit{``I am the kind of person who likes to see what I am focusing on... [Schemex] was a little too much for me.''}

\subsection{Design Implications}

Our findings suggest that the interactive structure of Schemex, rather than its outputs alone, was central to participants' structural understanding. 
From these findings, we derive several implications for designing interactive systems that support abstraction and discovery from examples.

\paragraph{Intermediate representations scaffold discovery.}
The clustering stage was not simply a preprocessing step; it was where many participants first recognized structural diversity in their data.
Systems that support abstraction from examples should consider surfacing intermediate groupings - even imperfect ones - as prompts for reflection, rather than jumping directly to a final structure.

\paragraph{Evidence grounding transforms trust dynamics.}
Participants' confidence was shaped less by the sophistication of the schema and more by their ability to verify it.
Inline citations, example matrices, and consistency indicators allowed users to evaluate proposals on their own terms.
For AI-assisted sensemaking tools, making the evidential basis of proposals inspectable may matter as much as improving proposal quality.

\paragraph{Contrastive generation externalizes tacit knowledge.}
The apply-compare-refine cycle gave participants a language for articulating what they knew implicitly.
Seeing where generated outputs diverged from originals made abstract gaps concrete.
This suggests that generation-based evaluation - where users assess AI outputs against known examples - can serve as a powerful elicitation technique, not just a quality metric.

\paragraph{Agency requires lightweight, localized controls.}
Participants engaged most productively when edits were small and reversible.
Systems that require users to specify complete abstractions upfront may discourage the exploratory, iterative engagement that makes discovery productive.
Supporting incremental revision preserves user agency without demanding comprehensive expertise.

\paragraph{Accommodating diverse information preferences.}
While most participants valued the layered, drill-down interface, some preferred more condensed representations.
Abstraction and sensemaking tools should consider offering multiple views - such as a compact summary alongside the full interactive canvas - to accommodate different cognitive styles and task contexts.
 
\section{Discussion}
\subsection{Supporting Schema Discovery through Interactive Structural Abstraction}
Schemex scaffolds structural abstraction through AI-assisted reasoning and visual interaction. Participants consistently preferred this workflow over chat-based interfaces and reported deeper insight -  unsurprising in retrospect. 
Cognitive theories like constructivism~\cite{fosnot2013constructivism} suggest people learn more effectively when they can manipulate and iterate over ideas, rather than passively skim AI-generated answers. By making structure visible and refinable, Schemex encourages more deliberate engagement.

Schema induction is rarely linear. Like other sensemaking processes, it involves generating hypotheses, testing them against data, and refining iteratively. Schemex manages this by scaffolding multiple levels of abstraction, from clusters to dimensions to attributes, mirroring how experts build understanding over time while enabling novices to follow a similar path. Although designed for creative and communicative tasks, this approach generalizes to other domains requiring structural abstraction from complex data, such as organizational process mining or user flow modeling.

\subsection{Contrastive Refinement as a Mechanism for Evolving Schema Understanding}
People learn by comparing examples - noticing variation, identifying what holds constant, and refining understanding through contrast. Schemex operationalizes this through contrastive refinement, comparing AI-generated outputs against real examples to surface mismatches and improve schemas iteratively, aligning with structural alignment theory~\cite{markman1993structural}. %Participants described the apply-compare-refine cycle as making tacit knowledge concrete: seeing where generated outputs diverged from originals gave them language for gaps they had sensed but could not articulate.

Because schemas are rarely static, this mechanism also helps users remain responsive as genres evolve. Contrastive refinement grounds schema revision in specific, observable differences rather than abstract intuition. Future work could extend this toward longitudinal schema tracking, examining how schemas shift within individuals or communities over time and how tools might help users navigate those shifts.

\subsection{Schemas as Scaffolding, Not Constraints: Toward Creative Agency}
An open question is what happens \textit{after} users internalize a schema. Schema theory suggests mastery involves not just acquiring a structure but knowing when to deviate from it~\cite{gick1983schema} - expert writers and designers often violate conventions deliberately to create meaning. How tools like Schemex might scaffold that transition from comprehension to creative deviation is a compelling future direction.

Preliminary evidence from a longitudinal study~\cite{wang2025role} suggests that schema understanding may function as a confidence threshold: once practitioners have a structural foothold, they feel less intimidated entering unfamiliar genres - but over time, they begin to break from the schema and develop their own voice. This arc - from imitation to internalization to departure - mirrors classical models of creative development and points to a key design challenge: how can schema-based tools support not just understanding, but eventual creative ownership?

\subsection{Limitations and Future Work}
Schemex converts images, videos, and audio into structured text to enable LLM-based reasoning across modalities, but this limits schema evaluation to a primarily textual modality. Future work could explore fully multi-modal comparison, better assessing how schemas function across media. The schema application workflow is also intentionally lightweight: given a user input and a derived schema, the system generates content by filling in values for each dimension. While sufficient for contrastive evaluation, real-world use raises open questions about input design, scaffolding, and guidance across different domains and expertise levels.

In our study, we used ChatGPT (o1-pro) as a baseline to give users a concrete point of comparison when reflecting on their Schemex experience - and it was the strongest publicly available tool at the time. More recently, agentic AI tools capable of autonomous multi-step iteration have emerged, and future work should study Schemex against these more capable baselines. That said, we believe the core affordances participants valued: evidence grounding, intermediate representations, and localized human control, reflect cognitive needs unlikely to be resolved by increased automation alone.

Finally, Schemex raises ethical concerns around scraping others' examples or imitating stylistic patterns without attribution~\cite{porquet2024copying}. Future versions should incorporate attribution tracking and usage transparency to encourage responsible use.
\section{Conclusion}
We introduced Schemex, an interactive workflow that helps users systematically uncover structural patterns from messy real-world examples. Through clustering, abstraction, and contrastive refinement, Schemex transforms implicit domain knowledge into explicit and actionable schemas. Our findings reveal that participants gained structural understanding not just from the schemas produced, but from the process of building them, pointing toward a broader design space for supporting structure discovery across diverse domains.

\bibliographystyle{ACM-Reference-Format}
\bibliography{sample-base,bib_vb}

%TC:ignore
\appendix
\section{Implementation}
\label{appendix:implementation}
The frontend is implemented in React, using React Flow to render the node-based workflow and side-panel interactions. The backend uses FastAPI for orchestrating model calls and Flask for persisting intermediate artifacts. We use o3-mini-high as the base model for its balance of speed and reasoning quality. For preprocessing, GPT-4V summarizes images and video frames, and Whisper transcribes audio. 
Prompts are provided in Appendix \ref{appendix:prompts_schemex}.

\section{Prompts used in Schemex}
\label{appendix:prompts_schemex}
\subsection{Get clusters}
\begin{Verbatim}[breaklines=true]
I’m learning how to {content_type}. 
Analyze the following examples to identify clusters based on STRUCTURAL and RHETORICAL patterns.
Focus on how the content is constructed and presented—not just what it’s about.
For textual examples, examine rhetorical organization.
For multimodal examples, consider how different modalities interact structurally and rhetorically.

For each cluster:
Name the structural approach (not just the topic)
List 2–3 shared structural features
Include the example IDs in that cluster

Clustering Rules:
Assign every example to exactly one cluster
Use all examples—no omissions or duplicates
Group based on structure, not content themes or ID order

Response Format:
Cluster 1: [Cluster Name]  
Common Features:  
- [Feature 1]  
- [Feature 2]  
Examples:  
- Example [ID]  
- Example [ID]  
...  
Total number of examples: [Number]  

Cluster 2: [Cluster Name]  
...

Examples to analyze: {examples}{input_context}
\end{Verbatim}

\subsection{Get the feature-example matrix}
\begin{Verbatim}[breaklines=true]
I’m learning how to {content_type}.
Help me analyze how each example in the cluster {cluster_name} demonstrates the identified common features.
Common Features: {common_features}
Examples: {examples}

Your Task:
For each example:
Indicate whether it demonstrates each feature:
"Yes" = Clearly demonstrates the feature
"Partial" = Feature is present but limited, implied, or modified
"No" = Feature is not present or is inconsistent with the definition

Include:
A brief explanation
A direct snippet from the example (if marked "Yes" or "Partial")

Format Requirements:
Return your output as a valid JSON object with the following structure:
{
  "mapping": [
    {
      "example_index": "EXAMPLE_ID",
      "example_snippet": "First 50 characters of the example...",
      "feature_mapping": [
        {
          "feature": "Feature 1",
          "feature_id": "F1",
          "applies": "Yes/No/Partial",
          "explanation": "Brief explanation of classification",
          "snippet": "Verbatim quote from example that demonstrates the feature"
        },
        ...
      ]
    },
    ...
  ]
}
\end{Verbatim}

\subsection{Infer dimensions and get the dimension-example matrix}
\begin{Verbatim}[breaklines=true]
I’m learning how to {user_goal}.
Analyze the following examples from the cluster: {cluster_name}
Examples: {examples_str}{input_context}

Your Task:
Help me infer the structural commonalities that define this cluster. Focus on identifying the key content dimensions that shape the narrative and composition.

Important Constraints — AVOID OVERFITTING:
Only identify dimensions shared across multiple examples
Focus on cluster-wide structural patterns, not individual quirks
Broaden dimensions only if needed to make them general enough for the cluster

For Each Dimension:
Give it a clear, descriptive name
Briefly explain what the dimension captures
Include verbatim snippets from examples that demonstrate it
For Each Example:
For every dimension, show:
Whether it applies: Yes, Partial, or No
A short explanation of why
A verbatim snippet from the example (if marked "Yes")
If no snippet can be found, use "Partial" or "No"
Snippet Rules:
Only use exact text from the examples—no paraphrasing or made-up text
If a snippet can’t be found, say so—never fabricate one

Applies Judgment:
Yes = Strong, specific snippet available
Partial = Present, but limited or implicit
No = Absent or not applicable

Output Format:
Return your response as a valid JSON object.
{
  "dimensions": [
    {
      "name": "Dimension name",
      "description": "Brief explanation",
      "examples": [
        {
          "example_id": "1",
          "snippet": "Verbatim snippet from Example 1"
        },
        ...
      ]
    },
    ...
  ],
  "example_mappings": [
    {
      "example_id": "1",
      "dimension_applications": [
        {
          "dimension": "Dimension name",
          "applies": "Yes/No/Partial",
          "explanation": "Brief explanation",
          "snippet": "Exact text from this example"
        },
        ...
      ]
    },
    ...
  ]
}
\end{Verbatim}

\subsection{Infer dimension-specific attributes and get the attribute-example matrix}
\begin{Verbatim}[breaklines=true]
I’m learning how to {user_goal}.
Please analyze the following examples from the cluster: {cluster_name}
Examples: {examples_full_text}{input_context}

PART 1: IDENTIFY DIMENSION ATTRIBUTES
You’ve already identified the following dimensions:
Dimensions: {dimensions_text}
Your task now is to define key attributes under each dimension, based on patterns that are consistent across the examples in this cluster.
Use only the following example IDs: Example IDs: {example_ids_text}

Attribute Requirements:
Each attribute must appear in at least 50% of the examples
Keep attributes broad enough to apply across examples, but still specific and observable
Exclude attributes that are too narrow, ambiguous, or inconsistently implemented
Evaluate every example against every attribute—no omissions

For Each Dimension, Provide:
Detailed Attributes: Clear, concrete, 1-sentence descriptions of each attribute
Concise Summaries: 1–2 word phrases summarizing each attribute (in the same order)

PART 2: ATTRIBUTE IMPLEMENTATION IN EXAMPLES
For every attribute in Part 1, assess how each example implements it.
Use the following classifications:
"YES" — Clearly and fully present (must include a direct quote)
"PARTIAL" — Present but limited, modified, or implicit
"NO" — Not present or structurally inconsistent with the attribute

Implementation Guidelines:
Always use the actual example IDs from the dataset
Provide a verbatim quote from the example if the classification is “YES” (never paraphrase)
Include an explanation for every example, even if “NO”
Never fabricate text—if you can’t find a quote, use "PARTIAL" or "NO"
Evaluate every example for every attribute, even if it’s “NO”

Critical Format Requirements (Strict JSON Only):
{
  "dimensions": {
    "Dimension Name": {
      "detailed": ["Detailed attribute 1", "Detailed attribute 2", ...],
      "concise": ["Summary 1", "Summary 2", ...]
    }
  },
  "attributes_examples": {
    "Dimension Name": {
      "Detailed attribute 1": [
        {
          "example_id": "ACTUAL_ID_1",
          "quote": "Exact quote from example",
          "explanation": "How this quote demonstrates the attribute",
          "classification": "YES"
        },
        {
          "example_id": "ACTUAL_ID_2",
          "quote": "",
          "explanation": "Why this example does not include this attribute",
          "classification": "NO"
        }
      ],
      ...
    }
  }
}
\end{Verbatim}

\subsection{Infer global constraints and get the attribute-example matrix}
\begin{Verbatim}[breaklines=true]
I’m learning how to {user_goal}.
Please analyze the following examples to identify overall structural patterns that are consistent across most of them.
Examples: {examples_full_text}{input_context}
Use only these example IDs in your analysis: Example IDs: {example_ids_text}

Goal: Identify Overall Attributes
Your task is to identify broad structural or compositional attributes that appear across the majority of examples, regardless of individual dimensions.
Look for consistent patterns in areas like:
Length – Word/sentence count range that appears in most examples
Format – Layout elements (e.g. paragraph count, headers, visual markers)
Tone – Overall voice, formality, or affective stance
Organization – Common sequence or structure across examples
Other global traits – Any other recurring patterns seen in most examples
You must include at least one attribute about length or format, and one about organization.

Attribute Guidelines
Each attribute must appear in at least 50% of the examples
Keep attributes broad enough to apply across many examples
Ground attributes in observable patterns (not interpretations)

For Each Attribute:
Provide a detailed description: a one-sentence explanation of the observable pattern
Provide a concise summary: a 1–2 word label for each attribute
The number of concise summaries must match the number of detailed attributes.

Attribute Application per Example
For each attribute, evaluate every example:
YES = Attribute is clearly and fully present (must include direct quote)
PARTIAL = Attribute is present but limited, modified, or implicit
NO = Attribute is absent or structurally inconsistent

For every example, include:
Verbatim quote from the example (if applicable)
A brief explanation of your classification
Classification: "YES", "PARTIAL", or "NO"

STRICT FORMAT (DO NOT DEVIATE):
{
  "overall_attributes": {
    "detailed": ["Detailed attribute 1", "Detailed attribute 2", "Detailed attribute 3"],
    "concise": ["Concise1", "Concise2", "Concise3"]
  },
  "overall_attributes_examples": {
    "Detailed attribute 1": [
      {
        "example_id": "ACTUAL_ID_1",
        "quote": "Exact quote from example",
        "explanation": "Why this example demonstrates the attribute",
        "classification": "YES"
      },
      {
        "example_id": "ACTUAL_ID_2",
        "quote": "",
        "explanation": "Why this example does not include the attribute",
        "classification": "NO"
      },
      ...
    ],
    "Detailed attribute 2": [...],
    "Detailed attribute 3": [...]
  }
}
\end{Verbatim}

\subsection{Get dimension values}
\begin{Verbatim}[breaklines=true]
I am trying to {current_user_goal}.
Given the following input: {input_text}
I need you to generate only one component of the output—not the full output.
Component to generate: {dim_name} – {dim_description}
Component Requirements: {attributes_text}

Important:
Focus only on generating the specified component.
Do not include other parts of the output.
Follow the given requirements closely.

\end{Verbatim}

\subsection{Get contrasting examples}
\begin{Verbatim}[breaklines=true]
I am trying to {current_user_goal}.
Given this input: {input_text}
And these dimension values: {dimensions_text}
Help me {current_user_goal} by generating a complete output.

Make sure to:
Integrate the dimension values effectively
Satisfy the following overall requirements: {overall_arrtibutes}
Output the final content directly. 

\end{Verbatim}

\subsection{Get comparative analysis}
\begin{Verbatim}[breaklines=true]
Your task is to analyze the gap between a generated output and a gold/reference example, in order to improve the schema itself.

Goal:
Your focus is on identifying what the schema is missing that would help the generated output better align with the gold example in more meaningful and sophisticated ways.

Inputs:
Schema: {schema_text}
Dimension Values: {dimension_values_text}
Generated Output: {generated_output}
Gold Example (Reference): {gold_example}

For Each Dimension:
Your analysis should address:
Patterns or qualities in the gold example that are not captured in the current schema
Gaps between the generated output and the gold example that a stronger schema could help bridge
Deeper or more nuanced attributes that could make the schema more generative, precise, and aligned with high-quality outputs

Think About:
What deeper structural or rhetorical qualities exist in the gold example that is missing from the schema?
Where is the schema too shallow, vague, or rigid?
How can the schema be refined or extended without overfitting?

For Each Dimension, Suggest 2–3 Improvements, Using these Tags:
[ADD] — Introduce a new attribute or pattern
[DEEPEN] — Make an existing attribute more sophisticated or layered
[REFINE] — Clarify or better define an existing attribute
[RESTRUCTURE] — Reorganize or reconceptualize the dimension
Also, include a section for Overall (cross-dimensional insights).

Output Format (STRICT):
Return your response as valid JSON using the following structure.
{
  "dimension_analysis": {
    "Dimension Name 1": {
      "analysis": "Analysis of deeper patterns in this dimension that aren't captured by the schema",
      "improvements": [
        "[ADD] Description of a new attribute",
        "[DEEPEN] Suggestion for making an existing attribute more sophisticated",
        "[REFINE] Clarify an attribute definition",
        "[RESTRUCTURE] Suggest restructuring if needed"
      ]
    },
    "Dimension Name 2": {
      "analysis": "Analysis of missing or underdefined qualities relevant to this dimension",
      "improvements": [
        "[ADD] ...",
        "[DEEPEN] ..."
      ]
    },
    "Overall": {
      "analysis": "Insights about patterns or improvements needed across dimensions",
      "improvements": [
        "[ADD] Cross-dimensional schema improvement",
        "[RESTRUCTURE] Suggest schema-wide restructuring"
      ]
    }
  }
}
\end{Verbatim}

\subsection{Get color coding}
\begin{Verbatim}[breaklines=true]
Your task is to create a detailed, structured comparison between two texts using schema dimensions to guide your analysis.
TEXTS TO COMPARE:
TEXT 1 (Generated Output): {generated_output}
TEXT 2 (Gold Example): {gold_example}
SCHEMA DIMENSIONS: {schema_text}

TASK:
Segment both texts and align corresponding parts side-by-side. For each segment:
Indicate which schema dimension (if any) the segment primarily reflects
Provide a brief annotation that explains the quality, relevance, or difference of this segment compared to its counterpart
Include the start and end character indices from the original source text
Set "dimension": null if the segment doesn’t clearly relate to any dimension
Rate the importance of this segment as "high", "medium", or "low" based on its role in conveying schema-aligned meaning or structure

CRITICAL REQUIREMENTS:
Cover the full content of both texts—no omissions allowed
Do not duplicate content across segments—each character should appear in exactly one segment
Accuracy of start_index and end_index is essential—these must correspond exactly to the characters in each original text
Use actual JSON null (not the string "null") when a segment doesn’t map to any dimension

OUTPUT FORMAT (Strict):
{
  "segments": [
    {
      "id": "segment_1",
      "source": "generated" | "gold",
      "text": "text of the segment",
      "start_index": 0,
      "end_index": 42,
      "dimension": "Dimension Name" | null,
      "annotation": "Brief analysis of the segment",
      "importance": "high" | "medium" | "low"
    },
    ...
  ],
  "dimension_analysis": {
    "Dimension Name 1": "Summary analysis of how this dimension compares across the two texts",
    "Dimension Name 2": "Additional dimension summary...",
    ...
  }
}
\end{Verbatim}

\subsection{Get iterated schema}
\begin{Verbatim}[breaklines=true]
You are helping improve a schema for AI-guided generation.
User Goal: {user_goal}
Context: {context_text}
The original schema includes:
Dimensions (each with detailed + concise attributes)
Overall attributes (for general output quality)

Your Task:
Use the suggestions below to iterate and improve the schema by:
Revising dimension attributes for clarity, depth, or generality
Adding new attributes where necessary
Improving overall attributes to enhance global output quality
Removing or simplifying overly specific attributes that may cause overfitting

Use These Inputs:
Suggested Improvements: {all_suggested_improvements}
Original Schema: {original_schema}

Critical Format Requirements:
For every dimension:
"detailed" and "concise" arrays must exist
Every "detailed" attribute must have a corresponding 
"concise" one, and vice versa
Follow the exact same JSON structure as the original schema

Final Output:
Return a valid JSON object with the same structure as the original schema, reflecting all suggested improvements.
\end{Verbatim}

\section{Prompts used for o1-pro in technical evaluation}
\label{appendix:prompts_o1_pro}
\begin{Verbatim}[breaklines=true]
I want to learn how to {content_type}. 

Here are {number_of_examples} examples. Can you help me infer the schema (structural commonalities) among the examples? So that I can use them later to create my own. You can give me multiple schemas if there are subgroups amongst the examples - each subgroup has one different schema. 

For each schema, please outline the key dimensions (characteristic features), their corresponding attributes (descriptors of the dimension), and overall attributes (general descriptors of the structure and content of the final output). 

Give me the answer in the format: for each schema, give it a descriptive name, list corresponding examples, overall attributes, dimensions, and attributes for dimension. 

Examples to analyze: {examples}{input_context}
\end{Verbatim}

%TC:endignore

\end{document}